# Diffusion Leaky Zero Attracting Least Mean Square Algorithm and Its Performance Analysis

**Long Shi**[1, 2]**, and Haiquan Zhao**[1, 2]**, Senior Member, IEEE**
[1]Key Laboratory of Magnetic Suspension Technology and Maglev Vehicle, Ministry of Education
[2] School of Electrical Engineering, Southwest Jiaotong University, Chengdu 610031, People's Republic of China

Corresponding author: Haiquan Zhao (e-mail: hqzhao_swjtu@126.com).

**ABSTRACT** Recently, the leaky diffusion least-mean-square (DLMS) algorithm has obtained much attention because of its good performance for high input eigenvalue spread and low signal-to-noise ratio (SNR). However, the leaky DLMS algorithm may suffer from performance deterioration in the sparse system. To overcome this drawback, the leaky zero attracting DLMS (LZA-DLMS) algorithm is developed in this paper, which adds an $l_1$-norm penalty to the cost function to exploit the property of sparse system. The leaky reweighted zero attracting DLMS (LRZA-DLMS) algorithm is also put forward, which can improve the estimation performance in the presence of time-varying sparsity. Instead of using the $l_1$-norm penalty, in the reweighted version, a log-sum function is employed as the substitution. Based on the weight error variance relation and several common assumptions, we analyze the transient behavior of our findings and determine the stability bound of the step-size. Moreover, we implement the steady state theoretical analysis for the proposed algorithms. Simulations in the context of distributed network system identification illustrate that the proposed schemes outperform various existing algorithms and validate the accuracy of the theoretical results.

## I. INTRODUCTION

PARAMETER estimation acts as an important role in the adaptive signal processing, which has obtained much attention over the past decades [1]-[8]. Recently, distributed estimation has increased popularity since it can deal with the information extraction from the data collected at nodes over the network, and also been applied in various fields, such as environment monitoring, disaster relief management and source localization [9]-[12]. In the previous work, two strategies have been extensively studied, namely the incremental strategy [13], [14] and the diffusion strategy [15], [16], respectively. In the incremental strategy, each node only communicates with its adjacent node in a sequential path. This strategy has low power requirements owing to simple communications [14], [17]. However, the incremental strategy is sensitive to link failure which is frequently encountered in the distributed network [18]-[20]. In contrast, the diffusion strategy is more widely applied in the distributed estimation due to its great robustness against the link failure [16], [21], [22]. In the diffusion strategy, each node over the network requires to communicate with all its neighbors and fuses the local estimates by a specific combination manner such as the uniform, Metropolis and relative-degree rules [23]-[25]. The implementation of diffusion strategy contains combination stage and adaptation stage. Based on different orders of these two stages, the *Combine-then-Adapt* (CTA) diffusion strategy and the *Adapt-then-Combine* (ATC) diffusion strategy were developed [22]. The previous literatures have illustrated that the ATC-type diffusion strategy outperforms the CTA-type strategy under a fair comparison [26]-[28].

In the family of diffusion algorithms, the diffusion least-mean-square (DLMS) algorithm was first proposed [15]. In [25], a more general diffusion algorithm was put forward, which allows measurement exchange in the adaptation stage. Subsequently, two modifications were developed to overcome the tradeoff between fast convergence rate and low steady state misalignment [29], [30]. In [31], the diffusion affine projection algorithm (DAPA) was proposed to speed up the convergence for colored inputs. To address other situations, various algorithms were investigated [20], [32]-[35].

In practical applications such as acoustic echo cancellation and active noise control, input signals usually exhibit the property of high eigenvalue spread and low signal-to-noise ratio (SNR) [36], [37], which can reduce the convergence rate and even result in the instability of the conventional DLMS algorithm [38]. To this end, the leaky DLMS algorithm was proposed [39], where the leakage term can prevent unbounded growth of the weight vectors from occurring so that the stability of the algorithm can be ensured [40]. Also note that in many practical fields, the unknown system is sparse (a large amount of coefficients are zeros or near-zeros) [41]-[43], which deteriorates the algorithm performance. It is reported that exploiting the sparsity of the unknown system is beneficial to enhance the estimation performance. Therefore, the zero attracting DLMS (ZA DLMS) and the reweighted zero attracting DLMS (RZA DLMS) were put forward, which can accelerate the convergence rate of near-zero coefficients in the sparse system [44], [45].

Motivated by the above facts, although the leaky DLMS algorithm equips superiority to deal with highly correlated inputs in a low SNR environment, we expect to further improve its estimation performance in the sparse system. Thus, this paper proposes the leaky zero attracting DLMS (LZA-DLMS) algorithm, which adds a penalty to the cost function to exploit the property of sparse system. As reported in [46], [47], the $l_0$-norm constraint renders the cost function non-convex, which leads to a Non-Polynomial (NP) hard problem on the minimization of the $l_0$-norm. Therefore, we employ the $l_1$-norm penalty for the proposed algorithms. In addition, the leaky reweighted zero attracting DLMS (LRZA-DLMS) algorithm is proposed for solving the time-varying sparsity problem. Unlike the LZA-DLMS algorithm, in the reweighted version, a log-sum function is employed as the constraint. Extensively, we develop their ATC and CTA variants, namely the ATC-LZA-DLMS, ATC-LRZA-DLMS, CTA-LZA-DLMS and CTA-LRZA-DLMS algorithms. In terms of theoretical contribution, we present the detailed transient analysis of our proposed ATC-type algorithms by invoking several assumptions, which can characterize the global weight error variance evolution. We also determine the stability bound of the step-size. In addition, we implement the steady state theoretical analysis for the ATC-type algorithms. Finally, Monte Carlo (MC) simulations conducted in various scenarios sufficiently demonstrate that the proposed algorithms outperform various existing algorithms and verify the accuracy of theoretical analysis.

The rest of the paper is organized as follows. In Section 2, we derive the LZA-DLMS and LRZA-DLMS algorithms including ATC and CTA versions. In Section 3, we analyze the transient and steady state behaviors of the ATC-type algorithms. We also discuss the stability bound of the step-size as well as the computational complexity. In Section 4, numerical simulations are conducted to test our findings and validate the theoretical analysis. In Section 5, we draw several conclusions.

*Notation*: We use normal letters to denote scalars and use boldface letters to denote vectors or matrices. The mathematical notations used in what follows are summarized in Table I.

TABLE I
MATHEMATICAL NOTATIONS

| Operators | Description |
|---|---|
| $(\cdot)^T$ | Transposition |
| $\otimes$ | Kronecker product |
| col$\{\cdot\}$ | Standard vectorization operation |
| vec$(\cdot)$ | Stack the columns of a matrix into a column vector |
| diag$\{\cdot\}$ | Diagonal matrix |
| $E[\cdot]$ | Expectation |
| Tr$(\cdot)$ | Trace operation |
| $\lambda_{\max}(\cdot)$ | Largest eigenvalue of a matrix |
| $\|\cdot\|$ | Absolute operation |
| $\|\cdot\|_p$ | $l_p$-norm |
| $\mathbf{I}_n$ | $n$ by $n$ identity matrix |

## II. PROPOSED ALGORITHMS

Consider a distributed network composed of $N$ nodes over a geographic region. At time $i$, each node $k$ has access to the time realization $\{d_k(i), \mathbf{u}_{k,i}\}$ of zero-mean random sequence. We are interested in estimating the unknown $M$-dimensional vector $\mathbf{w}_o$ that satisfies a linear model

$$d_k(i) = \mathbf{u}_{k,i}\mathbf{w}_o + v_k(i) \quad (1)$$

where $d_k(i)$ denotes the desired signal, $\mathbf{u}_{k,i} = [u_k(i), u_k(i-1), \cdots, u_k(i-M+1)]$ is the input vector, $\mathbf{w}_o = [w_o(1), w_o(2), \cdots, w_o(M)]^T$ stands for the unknown weight vector, and $v_k(i)$ represents the background noise with zero-mean and variance $\sigma_{v,k}^2$.

### A. THE LZA-DLMS ALGORITHM

In what follows, we consider the case of no measurement exchange in the adaptation stage [16]. Let $N_k$ denote the set of nodes in the neighborhood of node $k$ (including itself). To derive the ATC-LZA-DLMS algorithm, we can minimize the following cost function

$$J_k^{dist}(\mathbf{w}) = E\left|d_k(i) - \mathbf{u}_{k,i}\mathbf{w}\right|^2 + \sum_{l \in N_k/\{k\}} b_{l,k} \|\mathbf{w} - \boldsymbol{\varphi}_l\|^2 + \gamma \mathbf{w}^T \mathbf{w} + \rho \|\mathbf{w}\|_1 \quad (2)$$

where $b_{l,k}$ denotes the combination rule, $\gamma$ represents the positive leaky factor, $\rho$ is an attracting factor to balance the penalty and the estimation error, $\mathbf{w}$ and $\boldsymbol{\varphi}_l$ stand for the global and local estimates of $\mathbf{w}_o$, respectively. Note that the combination rule $b_{l,k}$ satisfies [22]

$$b_{l,k} = 0 \text{ if } l \notin N_k, \quad \mathbf{1}^T \boldsymbol{\Omega} = \mathbf{1}^T \quad (3)$$

where $\boldsymbol{\Omega}$ is the $N \times N$ matrix with individual entries $b_{l,k}$.

Taking the gradient of (2), we achieve

$$\left[\nabla_{\mathbf{w}} J_k^{dist}(\mathbf{w})\right] = \left(\mathbf{R}_{\mathbf{u},k}\mathbf{w} - \mathbf{R}_{d\mathbf{u},k}\right) + \sum_{l \in N_k/\{k\}} b_{l,k}(\mathbf{w} - \boldsymbol{\varphi}_l) + \gamma \mathbf{w} + \rho sign[\mathbf{w}] \quad (4)$$

where $\mathbf{R}_{\mathbf{u},k} = E[\mathbf{u}_{k,i}^T \mathbf{u}_{k,i}]$ is assumed positive-definite (i.e., $\mathbf{R}_{\mathbf{u},k} > 0$), and $\mathbf{R}_{d\mathbf{u},k} = E[d_k(i)\mathbf{u}_{k,i}^T]$ [15], [16].

Using the steepest-descent method for the estimate of $\mathbf{w}_o$ at node $k$, we obtain the following recursion

$$\mathbf{w}_{k,i} = \mathbf{w}_{k,i-1} + \mu\left(\mathbf{R}_{d\mathbf{u},k} - \mathbf{R}_{\mathbf{u},k}\mathbf{w}_{k,i-1}\right) - \mu\gamma\mathbf{w}_{k,i-1} - \rho sign[\mathbf{w}_{k,i-1}] + \chi \sum_{l \in N_k/\{k\}} b_{l,k}(\boldsymbol{\varphi}_l - \mathbf{w}_{k,i-1}) \quad (5)$$

where $\{\mu, \chi\}$ denote positive step-sizes.

**ATC-LZA-DLMS algorithm:** By introducing an intermediate $\boldsymbol{\varphi}_{k,i}$, Eq. (5) can be differentiated into two steps

$$\begin{cases} \boldsymbol{\varphi}_{k,i} = (1-\mu\gamma)\mathbf{w}_{k,i-1} + \mu\left(\mathbf{R}_{d\mathbf{u},k} - \mathbf{R}_{\mathbf{u},k}\mathbf{w}_{k,i-1}\right) - \rho sign[\mathbf{w}_{k,i-1}] \\ \mathbf{w}_{k,i} = \boldsymbol{\varphi}_{k,i} + \chi \sum_{l \in N_k/\{k\}} b_{l,k}(\boldsymbol{\varphi}_l - \mathbf{w}_{k,i-1}) \end{cases} \quad (6)$$

We then replace $\boldsymbol{\varphi}_l$ in (6) with the intermediate estimate $\boldsymbol{\varphi}_{l,i}$ which is available at node $l$ at time $i$, and also replace $\mathbf{w}_{k,i-1}$ with the intermediate estimate $\boldsymbol{\varphi}_{k,i}$. This substitution is reasonable since $\boldsymbol{\varphi}_{k,i}$ contains more information than $\mathbf{w}_{k,i-1}$ [39, 45], and leads to

$$\begin{cases} \boldsymbol{\varphi}_{k,i} = (1-\mu\gamma)\mathbf{w}_{k,i-1} + \mu\left(\mathbf{R}_{d\mathbf{u},k} - \mathbf{R}_{\mathbf{u},k}\mathbf{w}_{k,i-1}\right) - \rho sign[\mathbf{w}_{k,i-1}] \\ \mathbf{w}_{k,i} = \boldsymbol{\varphi}_{k,i} + \chi \sum_{l \in N_k/\{k\}} b_{l,k}(\boldsymbol{\varphi}_{l,i} - \boldsymbol{\varphi}_{k,i}) \end{cases} \quad (7)$$

Noting from the second equation in (7) and combining (3), we get

$$\mathbf{w}_{k,i} = (1 - \chi + \chi b_{k,k})\boldsymbol{\varphi}_{k,i} + \chi \sum_{l \in N_k/\{k\}} b_{l,k}\boldsymbol{\varphi}_{k,i} \quad (8)$$

As is done in [16], if we introduce the following coefficients

$$a_{k,k} = 1 - \chi + \chi b_{k,k} \text{ and } a_{k,k} = \chi b_{k,k} \text{ for } l \neq k \quad (9)$$

we have

$$\begin{cases} \boldsymbol{\varphi}_{k,i} = \mathbf{w}_{k,i-1} + \mu\left(\mathbf{R}_{d\mathbf{u},k} - \mathbf{R}_{\mathbf{u},k}\mathbf{w}_{k,i-1}\right) - \mu\gamma\mathbf{w}_{k,i-1} - \rho sign[\mathbf{w}_{k,i-1}] \\ \mathbf{w}_{k,i} = \sum_{l \in N_k} a_{l,k}\boldsymbol{\varphi}_{l,i} \end{cases} \quad (10)$$

where the coefficients $a_{l,k}$ are real, non-negative, and satisfy

$$a_{l,k} = 0 \text{ if } l \notin N_k, \quad \mathbf{1}^T\boldsymbol{\Gamma} = \mathbf{1}^T \quad (11)$$

where $\boldsymbol{\Gamma}$ is a $N \times N$ matrix with individual entries $a_{l,k}$.

Now employing the following instantaneous approximations for (10)

$$\mathbf{R}_{\mathbf{u},k} \approx \mathbf{u}_{k,i}^T \mathbf{u}_{k,i}, \qquad \mathbf{R}_{d\mathbf{u},k} \approx d_k(i)\mathbf{u}_{k,i}^T \quad (12)$$

we obtain the update equation for the proposed ATC-LZA-DLMS algorithm

$$\begin{cases} \boldsymbol{\varphi}_{k,i} = (1-\mu\gamma)\mathbf{w}_{k,i-1} + \mu\mathbf{u}_{k,i}^T(d_k(i) - \mathbf{u}_{k,i}\mathbf{w}_{k,i-1}) - \rho sign[\mathbf{w}_{k,i-1}] \\ \mathbf{w}_{k,i} = \sum_{l \in N_k} a_{l,k}\boldsymbol{\varphi}_{l,i} \end{cases} \quad (13)$$

**CTA-LZA-DLMS algorithm:** Note that we can also change the order of two steps in (6) and use a similar manner to implement derivation. Finally, we achieve the recursion for the CTA-LZA-DLMS algorithm

$$\begin{cases} \boldsymbol{\varphi}_{k,i} = \sum_{l \in N_k} a_{l,k}\mathbf{w}_{l,i-1} \\ \mathbf{w}_{k,i} = (1-\mu\gamma)\boldsymbol{\varphi}_{k,i} + \mu\mathbf{u}_{k,i}^T(d_k(i) - \mathbf{u}_{k,i}\boldsymbol{\varphi}_{k,i}) - \rho sign[\boldsymbol{\varphi}_{k,i}] \end{cases} \quad (14)$$

**Remark 1.** It can be observed from (13) and (14) that both the ATC-LZA-DLMS and CTA-LZA-DLMS algorithms have zero-attractors, denoted by $\rho sign[\mathbf{w}_{k,i-1}]$ and $\rho sign[\boldsymbol{\varphi}_{k,i}]$, which are functioned to shrink small weight coefficients to zeros in the sparse system, speeding up the convergence rate. However, if the system to be estimated is not sparse, the zero-attractors will also attract the non-sparse coefficients to zeros regardless of their amplitudes because the sign function only cares about the sign of coefficients, resulting in irrationality for large weight coefficients. Therefore, an improved version is presented below.

### B. THE LRZA-DLMS ALGORITHM

Motivated by the reweighted method [45], the LRZA-DLMS algorithm is derived by minimizing the following cost function

$$J_k^{dist}(\mathbf{w}) = E\left|d_k(i) - \mathbf{u}_{k,i}\mathbf{w}\right|^2 + \sum_{l \in N_k/\{k\}} b_{l,k}\left\|\mathbf{w} - \boldsymbol{\varphi}_l\right\|^2 + \gamma\mathbf{w}^T\mathbf{w} + \rho'\sum_{i=1}^{L}\log\left(1 + \frac{|w(i)|}{\varepsilon'}\right) \quad (15)$$

where $\rho'$ and $\varepsilon'$ are positive constants, and $w(i)$ denotes the $i$th entry of $\mathbf{w}$.

Since the derivation of the LRZA-DLMS algorithm is similar to that of the LZA-DLMS algorithm, we here omit its detailed process. We achieve the update equations for the ATC-LRZA-DLMS and CTA-LRZA-DLMS algorithms, as follows

**ATC-LRZA-DLMS algorithm:**

$$\begin{cases} \boldsymbol{\varphi}_{k,i} = (1-\mu\gamma)\mathbf{w}_{k,i-1} + \mu\mathbf{u}_{k,i}^T(d_k(i) - \mathbf{u}_{k,i}\mathbf{w}_{k,i-1}) - \rho\frac{sign[\mathbf{w}_{k,i-1}]}{1+\varepsilon|\mathbf{w}_{k,i-1}|} \\ \mathbf{w}_{k,i} = \sum_{l \in N_k} a_{l,k}\boldsymbol{\varphi}_{l,i} \end{cases} \quad (16)$$

and

**CTA-LRZA-DLMS algorithm:**

$$\begin{cases} \boldsymbol{\varphi}_{k,i} = \sum_{l \in N_k} a_{l,k} \mathbf{w}_{l,i-1} \\ \mathbf{w}_{k,i} = (1-\mu\gamma)\boldsymbol{\varphi}_{k,i} + \mu\mathbf{u}_{k,i}^T(d_k(i) - \mathbf{u}_{k,i}\boldsymbol{\varphi}_{k,i}) - \rho\frac{sign[\boldsymbol{\varphi}_{k,i}]}{1+\varepsilon|\boldsymbol{\varphi}_{k,i}|} \end{cases} \quad (17)$$

where $\rho = \mu\rho'/\varepsilon'$ and $\varepsilon = 1/\varepsilon'$.

**Remark 2.** The zero-attractors in (16) and (17) are denoted by $\frac{sign[\mathbf{w}_{k,i-1}]}{1+\varepsilon|\mathbf{w}_{k,i-1}|}$ and $\frac{sign[\boldsymbol{\varphi}_{k,i}]}{1+\varepsilon|\boldsymbol{\varphi}_{k,i}|}$, respectively. It can be seen that the zero-attractor not only shrinks small weight coefficients to zeros, but also distinguishes non-zero coefficients because it reflects the effect of amplitudes instead of directly taking the signs of the coefficients. For a better understanding, the proposed algorithms are summarized in Table II.

TABLE II
SUMMARY OF THE ALGORITHMS

| ATC-type | CTA-type |
|---|---|
| Initialization: $\mathbf{w}_{k,-1} = \boldsymbol{\varphi}_{k,-1} = 0$ for each node $k$ | Initialization: $\mathbf{w}_{k,-1} = \boldsymbol{\varphi}_{k,-1} = 0$ for each node $k$ |
| Adaptation by the ATC-type algorithms: ATC-LZA-DLMS $\boldsymbol{\varphi}_{k,i} = (1-\mu\gamma)\mathbf{w}_{k,i-1} + \mu\mathbf{u}_{k,i}^T(d_k(i) - \mathbf{u}_{k,i}\mathbf{w}_{k,i-1}) - \rho sign[\mathbf{w}_{k,i-1}]$ ATC-LRZA-DLMS $\boldsymbol{\varphi}_{k,i} = (1-\mu\gamma)\mathbf{w}_{k,i-1} + \mu\mathbf{u}_{k,i}^T(d_k(i) - \mathbf{u}_{k,i}\mathbf{w}_{k,i-1}) - \rho\frac{sign[\mathbf{w}_{k,i-1}]}{1+\varepsilon|\mathbf{w}_{k,i-1}|}$ Combination: $\mathbf{w}_{k,i} = \sum_{l \in N_k} a_{l,k}\boldsymbol{\varphi}_{l,i}$ | Combination: $\boldsymbol{\varphi}_{k,i} = \sum_{l \in N_k} a_{l,k}\mathbf{w}_{l,i-1}$ Adaptation by the CTA-type algorithms: CTA-LZA-DLMS $\mathbf{w}_{k,i} = (1-\mu\gamma)\boldsymbol{\varphi}_{k,i} + \mu\mathbf{u}_{k,i}^T(d_k(i) - \mathbf{u}_{k,i}\boldsymbol{\varphi}_{k,i}) - \rho sign[\boldsymbol{\varphi}_{k,i}]$ CTA-LRZA-DLMS $\mathbf{w}_{k,i} = (1-\mu\gamma)\boldsymbol{\varphi}_{k,i} + \mu\mathbf{u}_{k,i}^T(d_k(i) - \mathbf{u}_{k,i}\boldsymbol{\varphi}_{k,i}) - \rho\frac{sign[\boldsymbol{\varphi}_{k,i}]}{1+\varepsilon|\boldsymbol{\varphi}_{k,i}|}$ |

## III. PERFORMANCE ANALYSIS

In this section, we perform the transient behavior analysis and determine the stability bound of the step-size for the proposed algorithms. Moreover, we analyze the steady state performance, as well as discuss the computational complexity. Since the ATC and CTA algorithms have similarities in terms of the analysis, we only carry out the analysis for the ATC-type algorithms as a demonstration. In order to make analysis tractable, we utilize the following unified model to characterize the ATC-type algorithms

$$\begin{cases} \boldsymbol{\varphi}_{k,i} = (1-\mu\gamma)\mathbf{w}_{k,i-1} + \mu\mathbf{u}_{k,i}^T e_{k,i} - \rho g[\mathbf{w}_{k,i-1}] \\ \mathbf{w}_{k,i} = \sum_{l \in N_k} a_{l,k}\boldsymbol{\varphi}_{l,i} \end{cases} \quad (18)$$

where $g[\mathbf{w}_{k,i-1}]$ denotes $sign[\mathbf{w}_{k,i-1}]$ for the ATC-LZA-DLMS algorithm, and represents $\frac{sign[\mathbf{w}_{k,i-1}]}{1+\varepsilon|\mathbf{w}_{k,i-1}|}$ for the ATC-LRZA-DLMS algorithm. To proceed, it is necessary to introduce some statistical assumptions and approximations.

**Assumption 1.** The regressors $\mathbf{u}_{k,i}$ are temporally and spatially independent, and identically distributed (i.i.d.) with zero-mean [16], [25], [26].

**Assumption 2.** The background noise $v_k(i)$ is i.i.d. with zero-mean and variance $\sigma_{v_k}^2$, and is independent of $\mathbf{u}_{k,i}$ [9], [21].

**Assumption 3.** The $m$th entry of the weight error vector at node $k$ at time $i$, namely $\tilde{w}_{k,i}(m)$, is subject to Gaussian distribution with mean $\mu_{k,i}(m)$ and variance $\sigma_{k,i}^2(m)$, i.e., $\tilde{w}_{k,i}(m) \sim N(\mu_{k,i}(m), \sigma_{k,i}^2(m))$ [48]-[51]. Thus, the $m$th entry of the estimated weight vector $\mathbf{w}_{k,i}$ follows Gaussian distribution, expressed as

$$w_{k,i}(m) = w_o(m) - \tilde{w}_{k,i}(m) \sim N(w_o(m) - \mu_{k,i}(m), \sigma_{k,i}^2(m))$$

**Approximation 1.** For $m \neq n$, we make the approximations
$E(g(w_{k,i}(m))g(w_{k,i}(n))) \approx E(g(w_{k,i}(m)))E(g(w_{k,i}(n)))$ and
$E(w_{k,i}(m)g(w_{k,i}(n))) \approx E(w_{k,i}(m))E(g(w_{k,i}(n)))$ [48], [50], [51].

**Approximation 2.** The fluctuations of $w_{k,i}(m)$ from one iteration to the next iteration are small enough to achieve

$$E\left(\frac{sign[w_{k,i-1}(m)]}{1+\varepsilon|w_{k,i-1}(m)|}\right) \approx \frac{E(sign[w_{k,i-1}(m)])}{1+\varepsilon E|w_{k,i-1}(m)|},$$

$$E\left(\frac{|w_{k,i-1}(m)|}{1+\varepsilon|w_{k,i-1}(m)|}\right) \approx \frac{E|w_{k,i-1}(m)|}{1+\varepsilon E|w_{k,i-1}(m)|} \text{ and}$$

$$E\left(\frac{1}{\left(1+\varepsilon|w_{k,i-1}(m)|\right)^2}\right) \approx \frac{1}{E\left(1+\varepsilon|w_{k,i-1}(m)|\right)^2} \text{ [51], [52].}$$

**Remark 3.** Assumptions 1~3 have been successfully used in analyzing the adaptive filtering algorithms although these assumptions may not be true in practical applications.

Approximations 1~2 are beneficial to the calculation of expectations expressed by nonlinear functions of adaptive tap-weights, which has been verified as an effective methodology, especially in the case of white input signals [50], [51]. Furthermore, using these approximations makes it feasible to predict the behaviors of the proposed algorithms.

## A. MEAN BEHAVIOR MODEL

We define the weight error vectors $\tilde{\mathbf{w}}_{k,i} = \mathbf{w}_o - \mathbf{w}_{k,i}$, $\tilde{\boldsymbol{\varphi}}_{k,i} = \mathbf{w}_o - \boldsymbol{\varphi}_{k,i}$, and the global quantities of the network vectors and matrices

$$\mathbf{w}_{opt} \triangleq \mathrm{col}\{\mathbf{w}_o, \mathbf{w}_o, \cdots, \mathbf{w}_o\} \tag{19}$$

$$\mathbf{w}_i \triangleq \mathrm{col}\{\mathbf{w}_{1,i}, \mathbf{w}_{2,i}, \cdots, \mathbf{w}_{N,i}\} \tag{20}$$

$$\tilde{\mathbf{w}}_i \triangleq \mathrm{col}\{\tilde{\mathbf{w}}_{1,i}, \tilde{\mathbf{w}}_{2,i}, \cdots, \tilde{\mathbf{w}}_{N,i}\} \tag{21}$$

$$\tilde{\boldsymbol{\varphi}}_i \triangleq \mathrm{col}\{\tilde{\boldsymbol{\varphi}}_{1,i}, \tilde{\boldsymbol{\varphi}}_{2,i}, \cdots, \tilde{\boldsymbol{\varphi}}_{N,i}\} \tag{22}$$

$$g[\mathbf{w}_i] \triangleq \mathrm{col}\{g[\mathbf{w}_{1,i}], g[\mathbf{w}_{2,i}], \cdots, g[\mathbf{w}_{N,i-1}]\} \tag{23}$$

$$\mathbf{U}_i \triangleq \mathrm{diag}\{\mathbf{u}_{1,i}, \mathbf{u}_{2,i}, \cdots, \mathbf{u}_{N,i}\} \tag{24}$$

Define the error vector, the noise vector, and the desired vector of the network

$$\mathbf{e}_i \triangleq \mathrm{col}\{e_{1,i}, e_{2,i}, \cdots, e_{N,i}\} \tag{25}$$

$$\mathbf{v}_i \triangleq \mathrm{col}\{v_1(i), v_2(i), \cdots, v_N(i)\} \tag{26}$$

$$\mathbf{d}_i \triangleq \mathrm{col}\{d_1(i), d_2(i), \cdots, d_N(i)\} \tag{27}$$

Also, the diagonal matrices for collecting the step-sizes $\mu$, leaky factors $\gamma$ and attracting factors $\rho$ are given by

$$\begin{aligned}\mathbf{M} &\triangleq \mathrm{diag}\{\mu, \mu, \cdots, \mu\} \\ \boldsymbol{\gamma}_s &\triangleq \mathrm{diag}\{\gamma, \gamma, \cdots, \gamma\} \\ \boldsymbol{\rho}_s &\triangleq \mathrm{diag}\{\rho, \rho, \cdots, \rho\}\end{aligned} \tag{28}$$

Considering $\mathbf{e}_i = \mathbf{U}_i \tilde{\mathbf{w}}_{i-1} + \mathbf{v}_i$ and rewriting the adaptation stage in (18) into the error vectors of the network yields

$$\tilde{\boldsymbol{\varphi}}_i = (\mathbf{I}_{MN} - \mathbf{Q}\mathbf{U}_i^T \mathbf{U}_i)\tilde{\mathbf{w}}_{i-1} + \mathbf{Q}\boldsymbol{\gamma}\mathbf{w}_{i-1} - \mathbf{Q}\mathbf{U}_i^T \mathbf{v}_i + \boldsymbol{\rho}g[\mathbf{w}_{i-1}] \tag{29}$$

where $\mathbf{Q} = \mathbf{M} \otimes \mathbf{I}_M$, $\boldsymbol{\gamma} = \boldsymbol{\gamma}_s \otimes \mathbf{I}_M$, and $\boldsymbol{\rho} = \boldsymbol{\rho}_s \otimes \mathbf{I}_M$.

Taking into account the combination stage in (18), the recursion can be integrated into

$$\tilde{\mathbf{w}}_i = \mathbf{P}(\mathbf{I}_{MN} - \mathbf{Q}\mathbf{U}_i^T \mathbf{U}_i)\tilde{\mathbf{w}}_{i-1} + \mathbf{P}\mathbf{Q}\boldsymbol{\gamma}\mathbf{w}_{i-1} - \mathbf{P}\mathbf{Q}\mathbf{U}_i^T \mathbf{v}_i + \mathbf{P}\boldsymbol{\rho}g[\mathbf{w}_{i-1}] \tag{30}$$

where $\mathbf{P} = \boldsymbol{\Gamma} \otimes \mathbf{I}_M$.

Now, taking expectations of both sides of (30) and invoking Assumption 1 and Assumption 2, we obtain

$$E[\tilde{\mathbf{w}}_i] = \mathbf{P}(\mathbf{I}_{MN} - \mathbf{Q}E[\mathbf{U}_i^T \mathbf{U}_i])E[\tilde{\mathbf{w}}_{i-1}] + \mathbf{P}\mathbf{Q}\boldsymbol{\gamma}E[\mathbf{w}_{i-1}] + \mathbf{P}\boldsymbol{\rho}E(g[\mathbf{w}_{i-1}]) \tag{31}$$

where $E[\mathbf{U}_i^T \mathbf{U}_i] = \mathbf{S} \otimes \mathbf{I}_M$ with $\mathbf{S} \triangleq diag\{\sigma_{u_1}^2, \sigma_{u_2}^2, \cdots, \sigma_{u_N}^2\}$. The expectation $E(g[\mathbf{w}_{i-1}])$ can be calculated below.

**ATC-LZA-DLMS algorithm:** For the ATC-LZA-DLMS, $E(g[\mathbf{w}_{i-1}])$ is denoted by $E(sign[\mathbf{w}_{k,i-1}])$. The $m$th component of $E(sign[\mathbf{w}_{k,i-1}])$ is given by $E(sign[w_{k,i-1}(m)])$. Applying Assumption 3, we can calculate $E(sign[w_{k,i-1}(m)])$.

$$\begin{aligned}&E(sign[w_{k,i-1}(m)]) \\ &= -\int_{-\infty}^{0} \frac{1}{\sqrt{2\pi\sigma_{k,i-1}^2(m)}} \exp\left(-\frac{\left(x - \left(w_o(m) - \mu_{k,i-1}(m)\right)\right)^2}{2\sigma_{k,i-1}^2(m)}\right)dx \\ &\quad + \int_{0}^{+\infty} \frac{1}{\sqrt{2\pi\sigma_{k,i-1}^2(m)}} \exp\left(-\frac{\left(x - \left(w_o(m) - \mu_{k,i-1}(m)\right)\right)^2}{2\sigma_{k,i-1}^2(m)}\right)dx \\ &= 1 - 2\phi\left(-\frac{\left(w_o(m) - \mu_{k,i-1}(m)\right)}{\sigma_{k,i-1}(m)}\right) \\ &= -erf\left(\frac{-w_o(m) + \mu_{k,i-1}(m)}{\sqrt{2\sigma_{k,i-1}^2(m)}}\right)\end{aligned} \tag{32}$$

Where $\mu_{k,i-1}(m) = E[\tilde{w}_{k,i-1}(m)]$, $\sigma_{k,i-1}^2(m) = (\tilde{\mathbf{W}}_{k,i})_{m,m} - (E[\tilde{w}_{k,i}(m)])^2$ with $\tilde{\mathbf{W}}_i \triangleq E[\tilde{\mathbf{w}}_i \tilde{\mathbf{w}}_i^T]$, $\phi(\cdot)$ denotes the cumulative distribution function (CDF) of the standard normal distribution, and $erf(\cdot)$ is the error function which is defined as $erf(x) = \frac{2}{\sqrt{\pi}} \int_0^x \exp(-t^2)dt$.

**ATC-LRZA-DLMS algorithm:** For the ATC-LRZA-DLMS, $E(g[\mathbf{w}_{i-1}])$ is denoted by $E\left(\frac{sign[\mathbf{w}_{k,i-1}]}{1+\varepsilon|\mathbf{w}_{k,i-1}|}\right)$. The $m$th component of $E\left(\frac{sign[\mathbf{w}_{k,i-1}]}{1+\varepsilon|\mathbf{w}_{k,i-1}|}\right)$ is expressed as $E\left(\frac{sign[w_{k,i-1}(m)]}{1+\varepsilon|w_{k,i-1}(m)|}\right)$. Employing Approximation 2, we have $E\left(\frac{sign[w_{k,i-1}(m)]}{1+\varepsilon|w_{k,i-1}(m)|}\right) \approx \frac{E(sign[w_{k,i-1}(m)])}{1+\varepsilon E|w_{k,i-1}(m)|}$, and $E|w_{k,i-1}(m)|$ can be computed by invoking Assumption 3 [51]

$$E\left|w_{k,i-1}(m)\right| = (w_o(m) - \mu_{k,i-1}(m)) erf\left(\frac{w_o(m) - \mu_{k,i-1}(m)}{\sqrt{2\sigma_{k,i-1}^2(m)}}\right)$$
$$+ \sqrt{\frac{2}{\pi}}\sigma_{k,i-1}(m)\exp\left(-\frac{(w_o(m) - \mu_{k,i-1}(m))^2}{2\sigma_{k,i-1}^2(m)}\right) \quad (33)$$

### B. MEAN SQUARE BEHAVIOR MODEL

Multiplying (30) by $\tilde{\mathbf{w}}_i^T$ for both sides gives rise to

$\tilde{\mathbf{w}}_i \tilde{\mathbf{w}}_i^T =$
$\mathbf{P}(\mathbf{I}_{MN} - \mathbf{Q}\mathbf{U}_i^T \mathbf{U}_i)\tilde{\mathbf{w}}_{i-1}\tilde{\mathbf{w}}_{i-1}^T(\mathbf{I}_{MN} - \mathbf{U}_i^T \mathbf{U}_i \mathbf{Q}^T)\mathbf{P}^T$
$+ \mathbf{P}\mathbf{Q}\boldsymbol{\gamma}\mathbf{w}_{i-1}\mathbf{w}_{i-1}^T\boldsymbol{\gamma}^T\mathbf{Q}^T\mathbf{P}^T + \mathbf{P}\mathbf{Q}\mathbf{U}_i^T \mathbf{v}_i \mathbf{v}_i^T \mathbf{U}_i \mathbf{Q}^T\mathbf{P}^T$
$+ \mathbf{P}\boldsymbol{\rho}g[\mathbf{w}_{i-1}]g[\mathbf{w}_{i-1}^T]\boldsymbol{\rho}^T\mathbf{P}^T + \mathbf{P}(\mathbf{I}_{MN} - \mathbf{Q}\mathbf{U}_i^T \mathbf{U}_i)\tilde{\mathbf{w}}_{i-1}\mathbf{w}_{i-1}^T\boldsymbol{\gamma}^T\mathbf{Q}^T\mathbf{P}^T$
$- \mathbf{P}(\mathbf{I}_{MN} - \mathbf{Q}\mathbf{U}_i^T \mathbf{U}_i)\tilde{\mathbf{w}}_{i-1}\mathbf{v}_i^T \mathbf{U}_i \mathbf{Q}^T\mathbf{P}^T + \mathbf{P}(\mathbf{I}_{MN} - \mathbf{Q}\mathbf{U}_i^T \mathbf{U}_i)\tilde{\mathbf{w}}_{i-1}g[\mathbf{w}_{i-1}^T]\boldsymbol{\rho}^T\mathbf{P}^T$
$+ \mathbf{P}\mathbf{Q}\boldsymbol{\gamma}\mathbf{w}_{i-1}\tilde{\mathbf{w}}_{i-1}^T(\mathbf{I}_{MN} - \mathbf{U}_i^T \mathbf{U}_i \mathbf{Q}^T)\mathbf{P}^T - \mathbf{P}\mathbf{Q}\boldsymbol{\gamma}\mathbf{w}_{i-1}\mathbf{v}_i^T \mathbf{U}_i \mathbf{Q}^T\mathbf{P}^T$
$+ \mathbf{P}\mathbf{Q}\boldsymbol{\gamma}\mathbf{w}_{i-1}g[\mathbf{w}_{i-1}^T]\boldsymbol{\rho}^T\mathbf{P}^T - \mathbf{P}\mathbf{Q}\mathbf{U}_i^T \mathbf{v}_i \tilde{\mathbf{w}}_{i-1}^T(\mathbf{I}_{MN} - \mathbf{U}_i^T \mathbf{U}_i \mathbf{Q}^T)\mathbf{P}^T$
$- \mathbf{P}\mathbf{Q}\mathbf{U}_i^T \mathbf{v}_i \mathbf{w}_{i-1}^T \boldsymbol{\gamma}^T \mathbf{Q}^T\mathbf{P}^T - \mathbf{P}\mathbf{Q}\mathbf{U}_i^T \mathbf{v}_i g[\mathbf{w}_{i-1}^T]\boldsymbol{\rho}^T\mathbf{P}^T$
$+ \mathbf{P}\boldsymbol{\rho}g[\mathbf{w}_{i-1}]\tilde{\mathbf{w}}_{i-1}^T(\mathbf{I}_{MN} - \mathbf{U}_i^T \mathbf{U}_i \mathbf{Q}^T)\mathbf{P}^T + \mathbf{P}\boldsymbol{\rho}g[\mathbf{w}_{i-1}]\mathbf{w}_{i-1}^T \boldsymbol{\gamma}^T \mathbf{Q}^T\mathbf{P}^T$
$- \mathbf{P}\boldsymbol{\rho}g[\mathbf{w}_{i-1}]\mathbf{v}_i^T \mathbf{U}_i \mathbf{Q}^T\mathbf{P}^T$
(34)

Taking expectations of both sides of (34) and invoking Assumption 1 and Assumption 2 yields

$E[\tilde{\mathbf{w}}_i \tilde{\mathbf{w}}_i^T] =$
$\mathbf{P}(\mathbf{I}_{MN} - \mathbf{Q}E[\mathbf{U}_i^T \mathbf{U}_i])E[\tilde{\mathbf{w}}_{i-1}\tilde{\mathbf{w}}_{i-1}^T](\mathbf{I}_{MN} - E[\mathbf{U}_i^T \mathbf{U}_i]\mathbf{Q}^T)\mathbf{P}^T$
$+ \mathbf{P}\mathbf{Q}\boldsymbol{\gamma}E[\mathbf{w}_{i-1}\mathbf{w}_{i-1}^T]\boldsymbol{\gamma}^T\mathbf{Q}^T\mathbf{P}^T + \mathbf{P}\mathbf{Q}E[\mathbf{U}_i^T \mathbf{v}_i \mathbf{v}_i^T \mathbf{U}_i]\mathbf{Q}^T\mathbf{P}^T$
$+ \mathbf{P}\boldsymbol{\rho}E\{g[\mathbf{w}_{i-1}]g[\mathbf{w}_{i-1}^T]\}\boldsymbol{\rho}^T\mathbf{P}^T + \mathbf{P}(\mathbf{I}_{MN} - \mathbf{Q}E[\mathbf{U}_i^T \mathbf{U}_i])E[\tilde{\mathbf{w}}_{i-1}\mathbf{w}_{i-1}^T]\boldsymbol{\gamma}^T\mathbf{Q}^T\mathbf{P}^T$
$+ \mathbf{P}(\mathbf{I}_{MN} - \mathbf{Q}E[\mathbf{U}_i^T \mathbf{U}_i])E\{\tilde{\mathbf{w}}_{i-1}g[\mathbf{w}_{i-1}^T]\}\boldsymbol{\rho}^T\mathbf{P}^T$
$+ \mathbf{P}\mathbf{Q}\boldsymbol{\gamma}E[\mathbf{w}_{i-1}\tilde{\mathbf{w}}_{i-1}^T](\mathbf{I}_{MN} - E[\mathbf{U}_i^T \mathbf{U}_i]\mathbf{Q}^T)\mathbf{P}^T + \mathbf{P}\mathbf{Q}\boldsymbol{\gamma}E\{\mathbf{w}_{i-1}g[\mathbf{w}_{i-1}^T]\}\boldsymbol{\rho}^T\mathbf{P}^T$
$+ \mathbf{P}\boldsymbol{\rho}E\{g[\mathbf{w}_{i-1}]\tilde{\mathbf{w}}_{i-1}^T\}(\mathbf{I}_{MN} - E[\mathbf{U}_i^T \mathbf{U}_i]\mathbf{Q}^T)\mathbf{P}^T + \mathbf{P}\boldsymbol{\rho}E\{g[\mathbf{w}_{i-1}]\mathbf{w}_{i-1}^T\}\boldsymbol{\gamma}^T\mathbf{Q}^T\mathbf{P}^T$
(35)

To implement the following analysis, we introduce the Kronecker product operation and its property [14], [24], [25]. That is, for arbitrary matrices $\{X, Y, Z\}$ which are compatible in terms of dimensions, we have $\text{vec}(XYZ) = (Z^T \otimes X)\text{vec}(Y)$.

Applying the above operation for (35), we can easily achieve

$\text{vec}(\tilde{\mathbf{W}}_i) =$
$(\mathbf{A}^T \otimes \mathbf{A}^T)\text{vec}(\tilde{\mathbf{W}}_{i-1}) + (\mathbf{B}^T \otimes \mathbf{B}^T)\text{vec}(\mathbf{W}_{i-1})$
$+ (\mathbf{C}^T \otimes \mathbf{C}^T)\text{vec}(E[\mathbf{U}_i^T \mathbf{v}_i \mathbf{v}_i^T \mathbf{U}_i])$
$+ (\mathbf{D}^T \otimes \mathbf{D}^T)\text{vec}(E(g[\mathbf{w}_{i-1}]g[\mathbf{w}_{i-1}^T]))$
$+ (\mathbf{B}^T \otimes \mathbf{A}^T)\text{vec}(E[\tilde{\mathbf{w}}_{i-1}\mathbf{w}_{i-1}^T]) + (\mathbf{D}^T \otimes \mathbf{A}^T)\text{vec}(E(\tilde{\mathbf{w}}_{i-1}g[\mathbf{w}_{i-1}^T]))$
$+ (\mathbf{A}^T \otimes \mathbf{B}^T)\text{vec}(E[\mathbf{w}_{i-1}\tilde{\mathbf{w}}_{i-1}^T]) + (\mathbf{D}^T \otimes \mathbf{B}^T)\text{vec}(E(\mathbf{w}_{i-1}g[\mathbf{w}_{i-1}^T]))$
$+ (\mathbf{A}^T \otimes \mathbf{D}^T)\text{vec}(E(g[\mathbf{w}_{i-1}]\tilde{\mathbf{w}}_{i-1}^T))$
$+ (\mathbf{B}^T \otimes \mathbf{D}^T)\text{vec}(E(g[\mathbf{w}_{i-1}]\mathbf{w}_{i-1}^T))$
(36)

where $\mathbf{W}_i \triangleq E[\mathbf{w}_i \mathbf{w}_i^T]$, $\mathbf{A} = (\mathbf{I}_{MN} - E[\mathbf{U}_i^T \mathbf{U}_i]\mathbf{Q}^T)\mathbf{P}^T$, $\mathbf{B} = \boldsymbol{\gamma}^T \mathbf{Q}^T \mathbf{P}^T$, $\mathbf{C} = \mathbf{Q}^T \mathbf{P}^T$, $\mathbf{D} = \boldsymbol{\rho}^T \mathbf{P}^T$. Under Assumption 1 and Assumption 2, $E[\mathbf{U}_i^T \mathbf{v}_i \mathbf{v}_i^T \mathbf{U}_i]$ can be expressed as

$$E[\mathbf{U}_i^T \mathbf{v}_i \mathbf{v}_i^T \mathbf{U}_i] = \mathbf{G} \otimes \mathbf{I}_M \quad (37)$$

where $\mathbf{G} \triangleq \text{diag}\{\sigma_{u_1}^2 \sigma_{v_1}^2, \sigma_{u_2}^2 \sigma_{v_2}^2, \cdots, \sigma_{u_N}^2 \sigma_{v_N}^2\}$.

**Remark 4.** The network mean-square deviation (MSD) is defined as the average of every MSD at node $k$, i.e., $\text{MSD}_{net,i} = \frac{1}{N}\sum_{k=1}^{N}\text{MSD}_{k,i}$. Note that $\text{MSD}_{net,i} = \frac{1}{N}\text{Tr}(\tilde{\mathbf{W}}_i)$, one can obtain the recursion for $\text{MSD}_{net,i}$ from (31) and (36). Then, the focus is to calculate several expectations in (36), including $E(g[\mathbf{w}_{i-1}]g[\mathbf{w}_{i-1}^T])$, $E[\tilde{\mathbf{w}}_{i-1}\mathbf{w}_{i-1}^T]$, $E(\tilde{\mathbf{w}}_{i-1}g[\mathbf{w}_{i-1}^T])$ and $E(\mathbf{w}_{i-1}g[\mathbf{w}_{i-1}^T])$. Given that $E[\tilde{\mathbf{w}}_{i-1}\mathbf{w}_{i-1}^T]$ and $E(\tilde{\mathbf{w}}_{i-1}g[\mathbf{w}_{i-1}^T])$ can be rewritten as $E[\tilde{\mathbf{w}}_{i-1}]\mathbf{w}_{opt}^T - E[\tilde{\mathbf{w}}_{i-1}\tilde{\mathbf{w}}_{i-1}^T]$ and $\mathbf{w}_{opt}E(g[\mathbf{w}_{i-1}^T]) - E(\mathbf{w}_{i-1}g[\mathbf{w}_{i-1}^T])$, we only need to take into account the calculation for $E(g[\mathbf{w}_{i-1}]g[\mathbf{w}_{i-1}^T])$ and $E(\mathbf{w}_{i-1}g[\mathbf{w}_{i-1}^T])$ so that (36) can be implemented. The expectations $E(g[\mathbf{w}_{i-1}]g[\mathbf{w}_{i-1}^T])$ and $E(\mathbf{w}_{i-1}g[\mathbf{w}_{i-1}^T])$ can be calculated by using Approximation 1 and Approximation 2.

### C. STABILITY BOUND OF THE STEP-SIZE

To ensure the proposed ATC-type algorithms can converge in the mean and mean-square, the bound of the step-size will be discussed in this part. From the mean aspect, the Eq. (31) can be reformulated as

$E[\tilde{\mathbf{w}}_i] = \mathbf{P}(\mathbf{I}_{MN} - \mathbf{Q}E[\mathbf{U}_i^T \mathbf{U}_i])E[\tilde{\mathbf{w}}_{i-1}] + \mathbf{P}\mathbf{Q}\boldsymbol{\gamma}E[\mathbf{w}_{i-1}] + \mathbf{P}\boldsymbol{\rho}E(g[\mathbf{w}_{i-1}])$
$= \mathbf{P}(\mathbf{I}_{MN} - \mathbf{Q}E[\mathbf{U}_i^T \mathbf{U}_i])E[\tilde{\mathbf{w}}_{i-1}] + \mathbf{P}\mathbf{Q}\boldsymbol{\gamma}E[\mathbf{w}_o - \tilde{\mathbf{w}}_{i-1}] + \mathbf{P}\boldsymbol{\rho}E(g[\mathbf{w}_{i-1}])$
$= \{\mathbf{P}(\mathbf{I}_{MN} - \mathbf{Q}E[\mathbf{U}_i^T \mathbf{U}_i]) - \mathbf{P}\mathbf{Q}\boldsymbol{\gamma}\}E[\tilde{\mathbf{w}}_{i-1}] + \mathbf{P}\mathbf{Q}\boldsymbol{\gamma}\mathbf{w}_o + \mathbf{P}\boldsymbol{\rho}E(g[\mathbf{w}_{i-1}])$
$= \{\mathbf{P}(\mathbf{I}_{MN} - \mu E[\mathbf{U}_i^T \mathbf{U}_i] - \mu\boldsymbol{\gamma})\}E[\tilde{\mathbf{w}}_{i-1}] + \mu\mathbf{P}\boldsymbol{\gamma}\mathbf{w}_o + \mathbf{P}\boldsymbol{\rho}E(g[\mathbf{w}_{i-1}])$
(38)

Note that $E(g[\mathbf{w}_{i-1}])$ only characterizes the zero-attractor, which has been proved to be bounded in the prior work [51], [52]. Therefore, the proposed algorithms will converge in

VOLUME XX, 2017

the mean if the condition $|\lambda_{\max}(\mathbf{PF})| < 1$ holds, where $\mathbf{F} \triangleq \mathbf{I}_{MN} - \mu E[\mathbf{U}_i^T \mathbf{U}_i] - \mu \boldsymbol{\gamma}$. Recalling that $\mathbf{P} = \boldsymbol{\Gamma} \otimes \mathbf{I}_M$, we get

$$|\lambda_{\max}(\mathbf{PF})| \leq \|\boldsymbol{\Gamma}\|_2 \cdot |\lambda_{\max}(\mathbf{F})| \tag{39}$$

Owing to the property of combination rule, $\|\boldsymbol{\Gamma}\|_2 \leq 1$ is guaranteed. Thus, the network stability in the mean depends on

$$|\lambda_{\max}(\mathbf{PF})| \leq |\lambda_{\max}(\mathbf{F})| \tag{40}$$

By deducing from (40), the ATC-type algorithms asymptotically converge in the mean if the step-size is chosen to satisfy

$$0 < \mu < \frac{2 - \gamma}{\lambda_{\max}(E[\mathbf{U}_i^T \mathbf{U}_i])} \tag{41}$$

We now consider the bound in the mean-square. To proceed, we rewrite (36) into an alternative formulation

$$\begin{aligned}
\text{vec}(\tilde{\mathbf{W}}_i) &= (\mathbf{A}^T \otimes \mathbf{A}^T + \mathbf{B}^T \otimes \mathbf{B}^T - \mathbf{B}^T \otimes \mathbf{A}^T - \mathbf{A}^T \otimes \mathbf{B}^T) \text{vec}(\tilde{\mathbf{W}}_{i-1}) \\
&+ (\mathbf{B}^T \otimes \mathbf{B}^T) \text{vec}(\mathbf{W}_o - \mathbf{w}_o E[\tilde{\mathbf{w}}_{i-1}^T] - E[\tilde{\mathbf{w}}_{i-1}] \mathbf{w}_o^T) \\
&+ (\mathbf{C}^T \otimes \mathbf{C}^T) \text{vec}(E[\mathbf{U}_i^T \mathbf{v}_i \mathbf{v}_i^T \mathbf{U}_i]) \\
&+ (\mathbf{D}^T \otimes \mathbf{D}^T) \text{vec}(E(g[\mathbf{w}_{i-1}] g[\mathbf{w}_{i-1}^T])) \\
&+ (\mathbf{B}^T \otimes \mathbf{A}^T) \text{vec}(E[\tilde{\mathbf{w}}_{i-1}] \mathbf{w}_o^T) \\
&+ (\mathbf{D}^T \otimes \mathbf{A}^T) \text{vec}(\mathbf{w}_o E(g[\mathbf{w}_{i-1}^T]) - E(\mathbf{w}_{i-1} g[\mathbf{w}_{i-1}^T])) \\
&+ (\mathbf{A}^T \otimes \mathbf{B}^T) \text{vec}(\mathbf{w}_o E[\tilde{\mathbf{w}}_{i-1}^T]) \\
&+ (\mathbf{D}^T \otimes \mathbf{B}^T) \text{vec}(E(\mathbf{w}_{i-1} g[\mathbf{w}_{i-1}^T])) \\
&+ (\mathbf{A}^T \otimes \mathbf{D}^T) \text{vec}(E(g[\mathbf{w}_{i-1}]) \mathbf{w}_o^T - E(g[\mathbf{w}_{i-1}] \mathbf{w}_{i-1}^T)) \\
&+ (\mathbf{B}^T \otimes \mathbf{D}^T) \text{vec}(E(g[\mathbf{w}_{i-1}] \mathbf{w}_{i-1}^T))
\end{aligned} \tag{42}$$

where $\mathbf{W}_o \triangleq \mathbf{w}_{opt} \mathbf{w}_{opt}^T$.

Also learn from [51], [52], the quantities $E(g[\mathbf{w}_{i-1}] g[\mathbf{w}_{i-1}^T])$, $E(\mathbf{w}_{i-1} g[\mathbf{w}_{i-1}^T])$, and $E(g[\mathbf{w}_{i-1}] \mathbf{w}_{i-1}^T)$ in (42) are bounded. Using the Kronecker product property $(XY) \otimes (ZW) = (X \otimes Z)(Y \otimes W)$ that is available for arbitrary matrices $\{X, Y, Z, W\}$ of compatible dimensions [27], we can express the terms $\mathbf{A}^T \otimes \mathbf{A}^T + \mathbf{B}^T \otimes \mathbf{B}^T - \mathbf{B}^T \otimes \mathbf{A}^T - \mathbf{A}^T \otimes \mathbf{B}^T$ in (42) as

$$\begin{aligned}
&\mathbf{A}^T \otimes \mathbf{A}^T + \mathbf{B}^T \otimes \mathbf{B}^T - \mathbf{B}^T \otimes \mathbf{A}^T - \mathbf{A}^T \otimes \mathbf{B}^T \\
&= (\mathbf{P} \otimes \mathbf{P}) \{\mathbf{I}_{M^2N^2} - \mu(\mathbf{I}_{MN} \otimes E[\mathbf{U}_i^T \mathbf{U}_i] + \mu E[\mathbf{U}_i^T \mathbf{U}_i] \otimes \mathbf{I}_{MN})\} \\
&+ \mu^2 (\mathbf{P} \otimes \mathbf{P}) \{E[(\mathbf{U}_i^T \mathbf{U}_i) \otimes (\mathbf{U}_i^T \mathbf{U}_i)]\} + \mu^2 (\mathbf{P} \otimes \mathbf{P})(\boldsymbol{\gamma} \otimes \boldsymbol{\gamma}) \\
&- \mu (\mathbf{P} \otimes \mathbf{P})(\boldsymbol{\gamma} \otimes \mathbf{I}_{MN}) + \mu^2 (\mathbf{P} \otimes \mathbf{P})(\boldsymbol{\gamma} \otimes E[\mathbf{U}_i^T \mathbf{U}_i]) \\
&- \mu (\mathbf{P} \otimes \mathbf{P})(\mathbf{I}_{MN} \otimes \boldsymbol{\gamma}) + \mu^2 (\mathbf{P} \otimes \mathbf{P})(E[\mathbf{U}_i^T \mathbf{U}_i] \otimes \boldsymbol{\gamma}) \\
&= (\mathbf{P} \otimes \mathbf{P}) \{\mu^2 \mathbf{K} - \mu \mathbf{J} + \mathbf{I}_{M^2N^2}\}
\end{aligned} \tag{43}$$

where

$$\begin{aligned}
\mathbf{K} &= E[(\mathbf{U}_i^T \mathbf{U}_i) \otimes (\mathbf{U}_i^T \mathbf{U}_i)] + \boldsymbol{\gamma} \otimes \boldsymbol{\gamma} + \boldsymbol{\gamma} \otimes E[\mathbf{U}_i^T \mathbf{U}_i] \\
&+ E[\mathbf{U}_i^T \mathbf{U}_i] \otimes \boldsymbol{\gamma}
\end{aligned} \tag{44}$$

and

$$\mathbf{J} = \mathbf{I}_{MN} \otimes E[\mathbf{U}_i^T \mathbf{U}_i] + E[\mathbf{U}_i^T \mathbf{U}_i] \otimes \mathbf{I}_{MN} + \boldsymbol{\gamma} \otimes \mathbf{I}_{MN} + \mathbf{I}_{MN} \otimes \boldsymbol{\gamma} \tag{45}$$

Therefore, the Eq. (42) can converge in the mean-square if $(\mathbf{P} \otimes \mathbf{P})\{\mu^2 \mathbf{K} - \mu \mathbf{J} + \mathbf{I}_{M^2N^2}\}$ is guaranteed to be stable. Due to the property of combination rule, we can ensure $\|\mathbf{P} \otimes \mathbf{P}\|_2 \leq 1$. It suggests that the proposed ATC-type algorithms can converge in the mean-square if $\mathbf{L} = \mu^2 \mathbf{K} - \mu \mathbf{J} + \mathbf{I}_{M^2N^2}$ is stable. Following the same argument in [51], the stable condition of matrix $\mathbf{L}$ can be determined

$$0 < \mu < \min\left\{\frac{1}{\lambda_{\max}(\mathbf{J}^{-1}\mathbf{K})}, \frac{1}{\max \lambda(\mathbf{H}) \in R^+}\right\} \tag{46}$$

where $\mathbf{H} = \begin{bmatrix} \frac{\mathbf{J}}{2} & -\frac{\mathbf{K}}{2} \\ \mathbf{I} & \mathbf{0} \end{bmatrix}$.

Therefore, to guarantee the stability of our ATC-type algorithms in the mean and mean-square, a stringent condition for the step-size is

$$0 < \mu < \min\left\{\frac{2-\gamma}{\lambda_{\max}(E[\mathbf{U}_i^T \mathbf{U}_i])}, \frac{1}{\lambda_{\max}(\mathbf{J}^{-1}\mathbf{K})}, \frac{1}{\max \lambda(\mathbf{H}) \in R^+}\right\} \tag{47}$$

### D. STEADY STATE PERFORMANCE

The network MSD in the steady state is defined by

$$\text{MSD}_{net,\infty} = \frac{1}{N} \text{Tr}(\tilde{\mathbf{W}}_\infty) \tag{48}$$

As $i \to \infty$, taking the limit and trace for (36), we obtain

$$\begin{aligned}
\text{Tr}(\tilde{\mathbf{W}}_\infty) &= \text{vec}(\mathbf{I}_{MN})^T \boldsymbol{\Phi}(\mathbf{B}^T \otimes \mathbf{B}^T) \text{vec}(\mathbf{W}_o) \\
&+ \text{vec}(\mathbf{I}_{MN})^T \boldsymbol{\Phi}(\mathbf{B}^T \otimes \mathbf{B}^T) \text{vec}(E(\mathbf{w}_o \tilde{\mathbf{w}}_\infty^T)) \\
&+ \text{vec}(\mathbf{I}_{MN})^T \boldsymbol{\Phi}(\mathbf{B}^T \otimes \mathbf{B}^T) \text{vec}(E(\tilde{\mathbf{w}}_\infty \mathbf{w}_o^T)) \\
&+ \text{vec}(\mathbf{I}_{MN})^T \boldsymbol{\Phi}(\mathbf{C}^T \otimes \mathbf{C}^T) \text{vec}(E[\mathbf{U}_\infty^T \mathbf{v}_\infty \mathbf{v}_\infty^T \mathbf{U}_\infty]) \\
&+ \text{vec}(\mathbf{I}_{MN})^T \boldsymbol{\Phi}(\mathbf{D}^T \otimes \mathbf{D}^T) \text{vec}(E(g[\mathbf{w}_\infty] g[\mathbf{w}_\infty^T])) \\
&+ \text{vec}(\mathbf{I}_{MN})^T \boldsymbol{\Phi}(\mathbf{B}^T \otimes \mathbf{A}^T) \text{vec}(E[\tilde{\mathbf{w}}_\infty \mathbf{w}_o^T]) \\
&+ \text{vec}(\mathbf{I}_{MN})^T \boldsymbol{\Phi}(\mathbf{D}^T \otimes \mathbf{A}^T) \text{vec}(E(\tilde{\mathbf{w}}_\infty g[\mathbf{w}_\infty^T])) \\
&+ \text{vec}(\mathbf{I}_{MN})^T \boldsymbol{\Phi}(\mathbf{A}^T \otimes \mathbf{B}^T) \text{vec}(E[\mathbf{w}_o \tilde{\mathbf{w}}_\infty^T]) \\
&+ \text{vec}(\mathbf{I}_{MN})^T \boldsymbol{\Phi}(\mathbf{D}^T \otimes \mathbf{B}^T) \text{vec}(E(\mathbf{w}_\infty g[\mathbf{w}_\infty^T])) \\
&+ \text{vec}(\mathbf{I}_{MN})^T \boldsymbol{\Phi}(\mathbf{A}^T \otimes \mathbf{D}^T) \text{vec}(E(g[\mathbf{w}_\infty] \tilde{\mathbf{w}}_\infty^T)) \\
&+ \text{vec}(\mathbf{I}_{MN})^T \boldsymbol{\Phi}(\mathbf{B}^T \otimes \mathbf{D}^T) \text{vec}(E(g[\mathbf{w}_\infty] \mathbf{w}_\infty^T))
\end{aligned} \tag{49}$$

where $\boldsymbol{\Phi} = (\mathbf{I}_{M^2N^2} - \mathbf{A}^T \otimes \mathbf{A}^T - \mathbf{B}^T \otimes \mathbf{B}^T + \mathbf{B}^T \otimes \mathbf{A}^T + \mathbf{A}^T \otimes \mathbf{B}^T)^{-1}$.

Then, taking the limit as $i \to \infty$ for (31), we arrive at

$$E[\tilde{\mathbf{w}}_\infty] = (\mathbf{I}_{MN} - \mathbf{A}^T)^{-1}(\mathbf{PQ}\gamma E[\mathbf{w}_\infty] + \mathbf{P}\boldsymbol{\rho}E(g[\mathbf{w}_\infty])) \quad (50)$$

To achieve the analytical results from (49) and (50), we make some assumptions in the steady state.

$$E(g[\mathbf{w}_\infty]g[\mathbf{w}_\infty^T]) \approx g[\mathbf{w}_o]g[\mathbf{w}_o^T] \quad (51)$$

$$E(\tilde{\mathbf{w}}_\infty g[\mathbf{w}_\infty^T]) \approx E(\tilde{\mathbf{w}}_\infty)g[\mathbf{w}_o^T] \quad (52)$$

$$E(g[\mathbf{w}_\infty]\tilde{\mathbf{w}}_\infty^T) \approx g[\mathbf{w}_o]E(\tilde{\mathbf{w}}_\infty^T) \quad (53)$$

These assumptions are reasonable because $\mathbf{w}_\infty \approx \mathbf{w}_o$ holds in the steady state. Combing (49) and (50) into (48), one can obtain $\text{MSD}_{net,\infty}$.

**Remark 5.** The detailed derivation of (49) is given in Appendix A. The developed analysis can also be used to obtain the steady state behaviors for the ATC DLMS, ATC ZA DLMS, and ATC leaky DLMS algorithms. For example, when $\mathbf{D} = \mathbf{0}$ and $\boldsymbol{\rho} = \mathbf{0}$, the proposed algorithm reduces to the ATC leaky DLMS algorithm. In this case, $E[\tilde{\mathbf{w}}_\infty]$ and $\text{Tr}(\tilde{\mathbf{W}}_\infty)$ are given by

$$E[\tilde{\mathbf{w}}_\infty] = (\mathbf{I}_{MN} - \mathbf{A}^T)^{-1}(\mathbf{PQ}\gamma E[\mathbf{w}_\infty]) \quad (54)$$

and

$$\text{Tr}(\tilde{\mathbf{W}}_\infty) = \text{vec}(\mathbf{I}_{MN})^T \boldsymbol{\Phi}(\mathbf{B}^T \otimes \mathbf{B}^T)\text{vec}(\mathbf{W}_o)$$
$$+ \text{vec}(\mathbf{I}_{MN})^T \boldsymbol{\Phi}(\mathbf{B}^T \otimes \mathbf{B}^T)\text{vec}(E(\mathbf{w}_o\tilde{\mathbf{w}}_\infty^T))$$
$$+ \text{vec}(\mathbf{I}_{MN})^T \boldsymbol{\Phi}(\mathbf{B}^T \otimes \mathbf{B}^T)\text{vec}(E(\tilde{\mathbf{w}}_\infty\mathbf{w}_o^T))$$
$$+ \text{vec}(\mathbf{I}_{MN})^T \boldsymbol{\Phi}(\mathbf{B}^T \otimes \mathbf{A}^T)\text{vec}(E[\tilde{\mathbf{w}}_\infty\mathbf{w}_o^T])$$
$$+ \text{vec}(\mathbf{I}_{MN})^T \boldsymbol{\Phi}(\mathbf{C}^T \otimes \mathbf{C}^T)\text{vec}(E[\mathbf{U}_\infty^T\mathbf{v}_\infty\mathbf{v}_\infty^T\mathbf{U}_\infty])$$
$$+ \text{vec}(\mathbf{I}_{MN})^T \boldsymbol{\Phi}(\mathbf{A}^T \otimes \mathbf{B}^T)\text{vec}(E[\mathbf{w}_o\tilde{\mathbf{w}}_\infty^T])$$
$$(55)$$

### E. COMPUTATIONAL COMPLEXITY

Table III summarizes the computational complexity including multiplications, additions and memory requirement for various algorithms, where $n_k$ denotes the number of components of the neighborhood set $N_k$, and $P$ stands for projection orders of the ATC DAPA. As compared with the ATC DLMS algorithm, the ATC DAPA has a significant increase in the complexity, while the ATC leaky DLMS, ATC ZA DLMS and ATC RZA DLMS algorithms just have a moderate increase in the complexity. As compared with the ATC ZA DLMS and ATC RZA DLMS algorithms, the proposed ATC-LZA-DLMS and ATC-LRZA-DLMS algorithms are more computationally expensive because additional calculation for leaky term is needed.

TABLE III
COMPUTATIONAL COMPLEXITY FOR NODE $k$ PER ITERATION

| Algorithms | Multiplications | Additions | Memory words |
|---|---|---|---|
| ATC DLMS [15] | $2M + Mn_k + 2$ | $2M + (M-1)n_k - 1$ | $(n_k + 2)M + 5$ |
| ATC DAPA [31] | $2P^2M + 2PM + M + Mn_k$ | $2P^2M + PM - P^2 + (n_k - 1)M$ | $MP + P^2 + 2M + 2P + (M+1)n_k$ |
| ATC leaky DLMS [39] | $2M + Mn_k + 3$ | $2M + (M-1)n_k$ | $(n_k + 3)M + 7$ |
| ATC ZA DLMS [45] | $3M + Mn_k + 2$ | $3M + (M-1)n_k - 1$ | $(n_k + 3)M + 6$ |
| ATC RZA DLMS [45] | $4M + Mn_k + 2$ | $4M + (M-1)n_k - 1$ | $(n_k + 3)M + 7$ |
| ATC-LZA-DLMS | $4M + Mn_k + 3$ | $3M + (M-1)n_k$ | $(n_k + 3)M + 7$ |
| ATC-LRZA-DLMS | $5M + Mn_k + 3$ | $4M + (M-1)n_k$ | $(n_k + 3)M + 8$ |

## IV. SIMULATIONS

In this section, we conduct Monte Carlo (MC) simulations to test the estimation performance of the proposed algorithms and evaluate the accuracy of the theoretical analysis. The adaptive filter and the unknown system are assumed to have the same number of taps. In Section 4.1, we show the estimation performance of the leaky algorithms and proposed algorithms in a synthetic sparse system. In Section 4.2, we compare the proposed algorithms with various existing algorithms for colored inputs. In Section 4.3, we verify the theoretical results by extensive simulations. The performance of all tested algorithms is evaluated by $10\log_{10}\text{MSD}_{net,i}$. The uniform rule is used in the simulations, which is defined as $a_{l,k} = 1/|N_k|$ for all $l$ [16]. Except the theoretical verification, simulation results are the average of 100 independent trials.

### A. SYNTHETIC SPARSE SYSTEM

In this subsection, we consider a network containing 20 nodes, shown in Fig. 1. The unknown system $\mathbf{w}_o$ has $M = 64$ taps. Initially, only one coefficient of $\mathbf{w}_o$ is set to 1 with its position randomly selected while other coefficients are equal to 0, making the system highly sparse. After 3000 iterations, 16 coefficients are set to 1 with their positions randomly selected, making the system have a sparsity ratio of 16/64. After 6000 iterations, 32 coefficients are randomly set equal to 1, yielding a non-sparse system. Both the Gaussian inputs and colored inputs are used to examine the algorithms. The variances of the Gaussian inputs and background noises are depicted in Fig. 2. The colored inputs are generated by passing the Gaussian inputs through a first-order system $G(z) = 1/(1-0.7z^{-1})$.

As can be seen from Fig. 3, when the system is highly sparse, the proposed ATC-type algorithms outperform the corresponding CTA-type algorithms regardless of Gaussian inputs or colored inputs. From Fig. 3(a), when the system is highly sparse, the LRZA-DLMS algorithm yields lower steady-state misalignment than the LZA-DLMS algorithm for both ATC and CTA types. The CTA-LRZA-DLMS algorithm is superior to the ATC-LZA-DLMS algorithm. Moreover, the proposed algorithms behave better than the leaky DLMS algorithm. When the system is less sparse (sparsity ratio 16/64), the CTA-LZA-DLMS and ATC-LZA-DLMS algorithms perform almost the same, meanwhile the CTA-LRZA-DLMS and ATC-LRZA-DLMS algorithms also achieve similar performance. However, in this stage, the leaky DLMS algorithm provides lower steady-state misalignment than the proposed algorithms. When the system is only half sparse, the performance of the proposed algorithms further deteriorates. From Fig. 3(b), it is clear that the ATC-LRZA-DLMS algorithm achieves the best performance no matter in which stage. When the system is highly sparse, the proposed algorithms outperform the leaky DLMS algorithm. Interestingly, the ATC-LZA-DLMS algorithm and the CTA-LZRA-DLMS algorithm almost have the same performance. When the system is less sparse (sparsity ratio 16/64), the CTA-LRZA-DLMS algorithm is superior to the ATC-LZA-DLMS algorithm. Meanwhile, the leaky DLMS algorithm yields lower steady-state misalignment than the LZA-DLMS algorithm. When the system is only half sparse, the performance of all the tested algorithms does not change significantly.

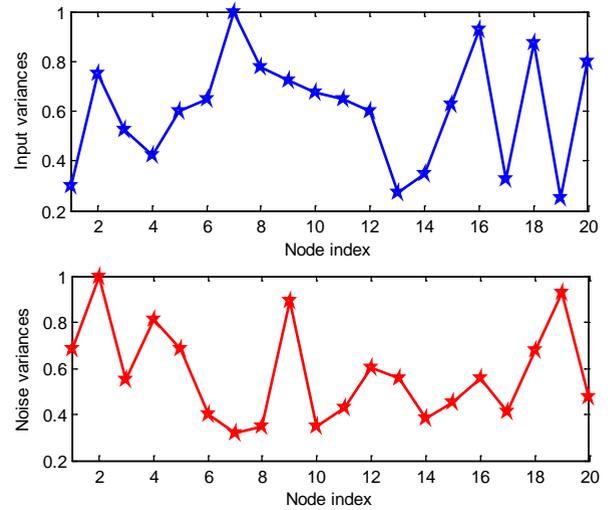

Fig. 2 Input variances and noise variances of 20 nodes

Therefore, the proposed algorithms are sensitive to the sparsity of the system. Fortunately, when the input is correlated, the ATC-LRZA-DLMS algorithm still maintains good performance even though the unknown system becomes non-sparse.

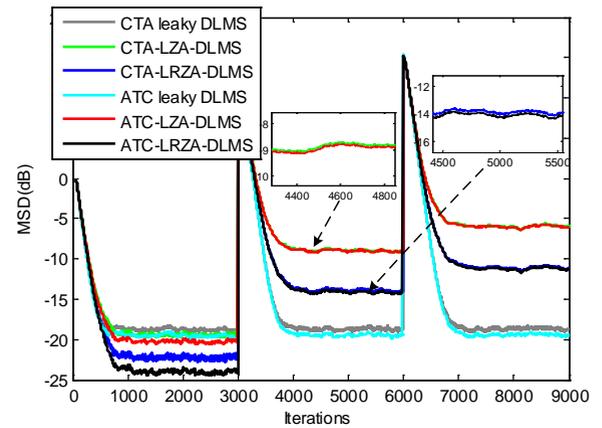

(a)

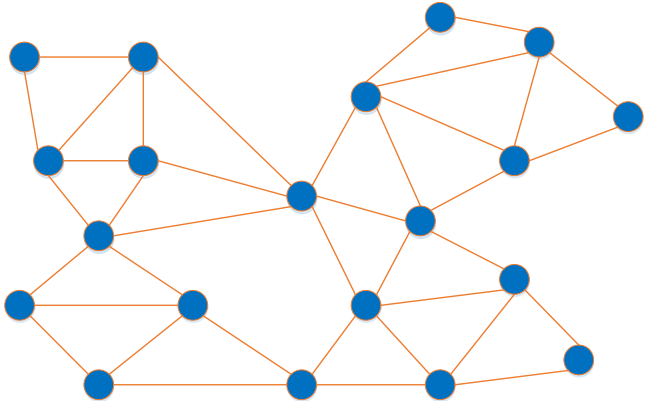

Fig. 1 Network topology of 20 nodes

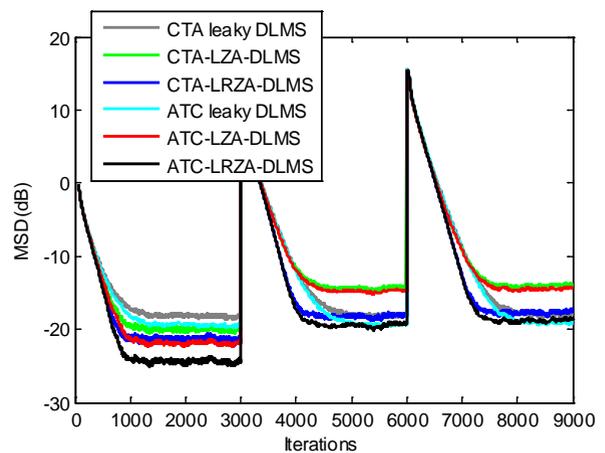

(b)

Fig. 3 MSD curves of the leaky DLMS and proposed algorithms. The leaky DLMS: $\mu$=0.01, and $\gamma$=0.002. The proposed algorithms: $\mu$=0.01,

$\gamma=0.002$, $\rho=0.0005$, and $\varepsilon=1$. (a) Gaussian inputs (b) colored inputs.

## B. SYSTEM IDENTIFICATION

In this subsection, we also employ the network shown in Fig. 1. The colored regressors $\mathbf{u}_{k,i}$ have length $M=128$, which are generated by filtering the Gaussian inputs through a first-order system $G(z)=1/(1-0.7z^{-1})$. The variances of the Gaussian inputs and background noises are the same as that in Section 4.1. In the simulations, the input signals firstly pass a highly sparse system whose coefficients are 0 except its first coefficient set to 1, then pass a system modeled by a FIR filter, whose frequency response and impulse response are depicted in Fig. 4. To facilitate the comparison, the proposed CTA-type algorithms are presented in Fig. 5 (a), and the ATC-type algorithms are shown in Fig.5 (b).

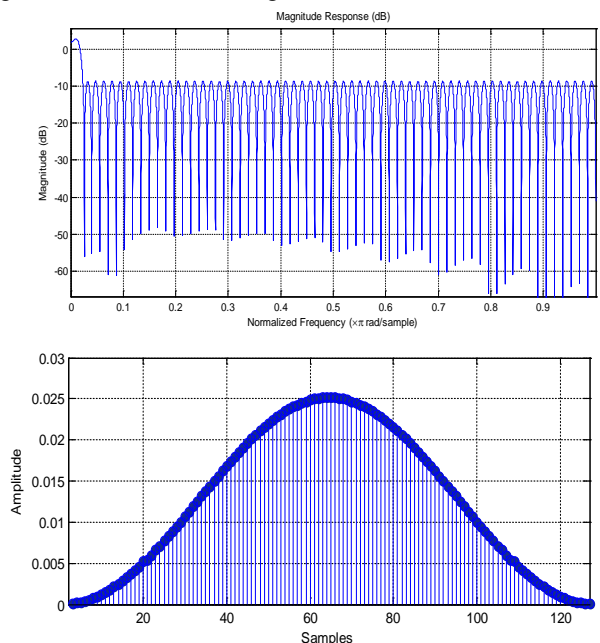

Fig. 4 Frequency response and impulse response of acoustic path

As can be seen from Fig. 5 (a), the CTA DLMS algorithm behaves worse than the CTA DAPA due to the effect of the colored inputs. In addition, both the CTA leaky DLMS and CTA ZA DLMS algorithms outperform the CTA DLMS algorithm thanks to the leaky factor and zero-attractor. It also suggests that the CTA RZA DLMS algorithm outperforms the CTA ZA DLMS algorithm because of the reweighted regularization. As compared with other tested algorithms, both the CTA-LZA-DLMS and CTA-LRZA-DLMS algorithms yield lower steady-state misalignment under the same convergence rate. In particular, the proposed CTA-LZA-DLMS is slightly inferior to the CTA-LRZA-DLMS algorithm in terms of the steady-state misalignment.

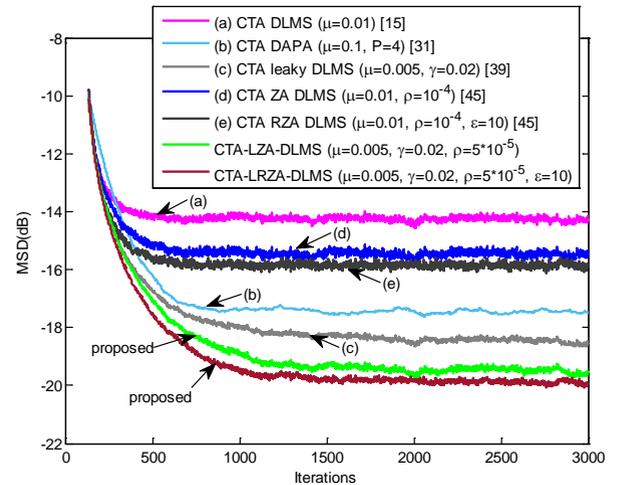

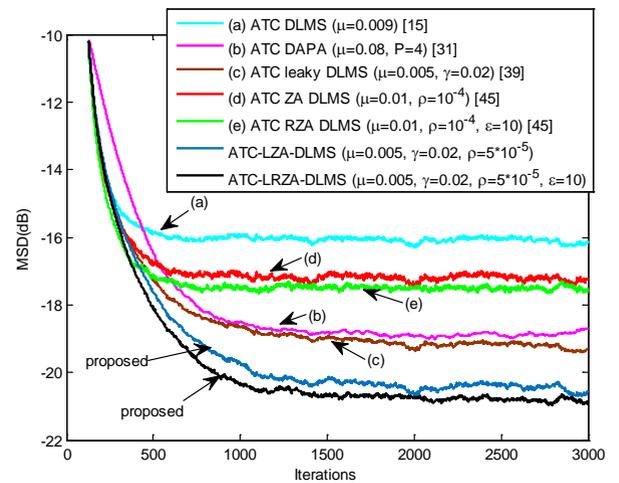

Fig. 5. The MSD curves of the proposed algorithms and some existing algorithms. (a) CTA-type (b) ATC-type.

As shown in Fig. 5 (b), the ATC DLMS algorithm still exhibits the poorest performance among all the algorithms, while the ATC leaky DLMS, ATC ZA DLMS and ATC RZA DLMS algorithms perform better than it. By contrast, the proposed algorithms are clearly superior to other algorithms, yielding much lower steady-state misalignment.

## C. TRANSIENT THEORETICAL VALIDATION

In this subsection, we verify the transient theoretical analysis for the proposed ATC-type algorithms. We consider a network composed of 5 nodes. The unknown system is modeled by $\mathbf{w}_o=[0\ 0\ 1\ 0\ 0]^T$. The variances of the Gaussian inputs and background noises are depicted in Fig. 6. The transient MSD curves are obtained from (31) and (36).

We firstly carry out the verification for the ATC-LZA-DLMS algorithm with respect to $\mu$. The parameters are selected as $\rho=0.005$, $\gamma=0.001$. As can be seen from Fig. 7, the theoretical results match well with the experimental results. Besides, it is found that the performance of the

VOLUME XX, 2017    9

ATC-LZA-DLMS algorithm is becoming better as the step-size $\mu$ increases. According to the characteristics of leaky algorithms and zero-attracting algorithms, it is reasonably inferred that the algorithm performance will deteriorate and be even unstable when the step-size increases to a certain value [53], [51].

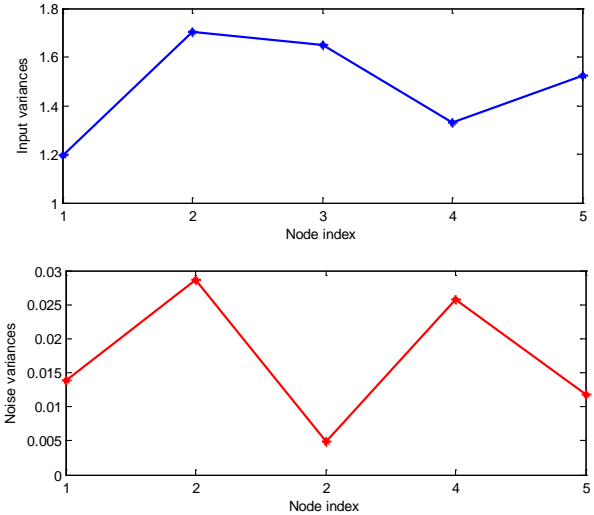

Fig. 6 Input variances and noise variances of 5 nodes

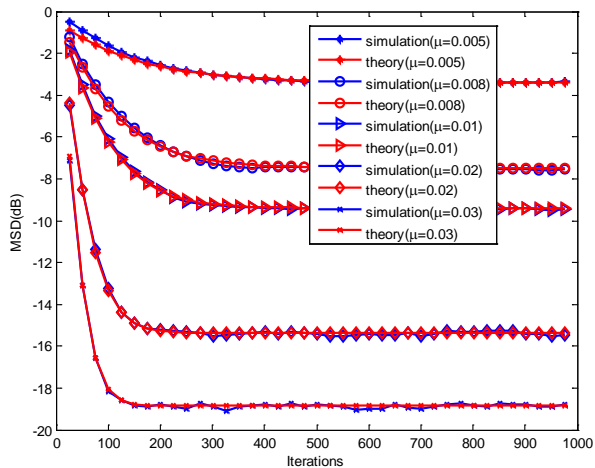

Fig. 7 The experimental and theoretical MSD curves of the proposed ATC-LZA-DLMS algorithm with respect to $\mu$

Then, we conduct the comparison for the ATC-LZA-DLMS algorithm with respect to $\gamma$, shown in Fig. 8. The parameters are set to $\mu = 0.03$, $\rho = 0.001$. As can be seen, the theoretical results also match accurately with the experimental results. It is observed that the ATC-LZA-DLMS algorithm with $\gamma = 0.001$ outperforms that with $\gamma = 0.01$ and $\gamma = 0.1$. Moreover, we implement the verification for the ATC-LZA-DLMS algorithm with respect to $\rho$, depicted in Fig. 9. The parameters are chosen as $\mu = 0.03$, $\gamma = 0.001$. It is obvious that the parameter $\rho$ makes a significant influence to the performance of the algorithms. For example, the ATC-LZA-DLMS algorithm with $\rho = 0.001$ is about 10dB lower than that with $\rho = 0.003$ in terms of the steady-state misalignment.

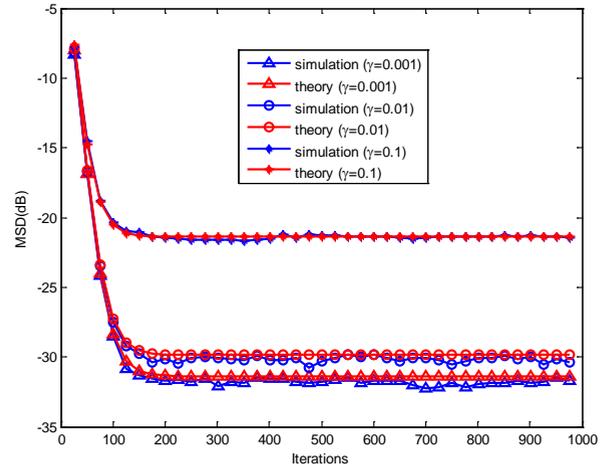

Fig. 8 The experimental and theoretical MSD curves of the proposed ATC-LZA-DLMS algorithm with respect to $\gamma$

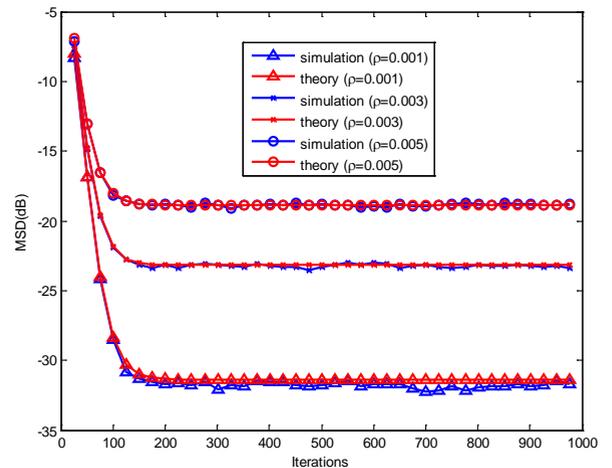

Fig. 9 The experimental and theoretical MSD curves of the proposed ATC-LZA-DLMS algorithm with respect to $\rho$

Next, we examine the theoretical accuracy for the ATC-LRZA-DLMS algorithm with respect to $\mu$. The parameters are selected as $\rho = 0.005$, $\gamma = 0.001$, $\varepsilon = 1$. As can be seen from Fig. 10, we can draw some conclusions similar to that in the ATC-LZA-DLMS algorithm. Furthermore, we conduct the comparison for the ATC-LRZA-DLMS algorithm with respect to $\gamma$, depicted in Fig. 11. The parameters are chosen as $\mu = 0.008$, $\rho = 0.001$, $\varepsilon = 1$. As can be seen, there is a small gap between the ATC-LRZA-DLMS algorithm with $\gamma = 0.001$ and that with $\gamma = 0.01$ in the steady-state misalignment. However, as the leaky factor $\gamma$ increases to 0.1, the performance of the algorithm rapidly deteriorates, about 8dB higher than that with $\gamma = 0.01$ in terms of the steady-state misalignment.

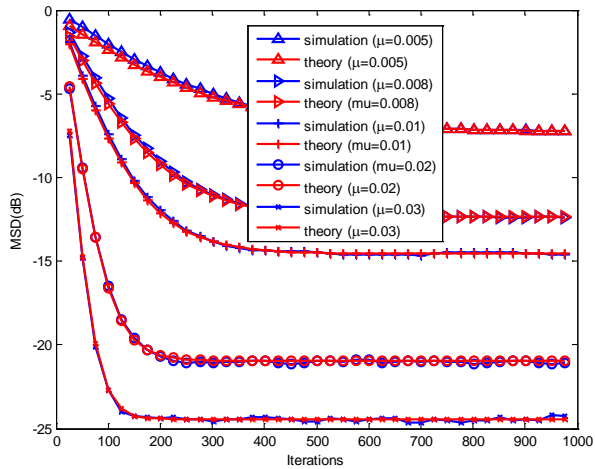

Fig. 10 The experimental and theoretical MSD curves of the proposed ATC-LRZA-DLMS algorithm with respect to $\mu$.

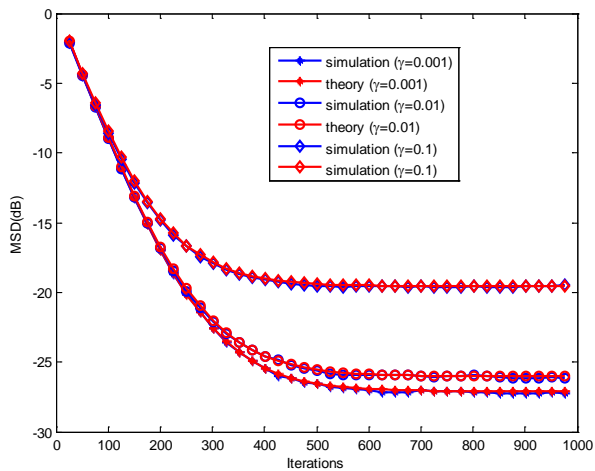

Fig. 11 The experimental and theoretical MSD curves of the proposed ATC-LRZA-DLMS algorithm with respect to $\gamma$.

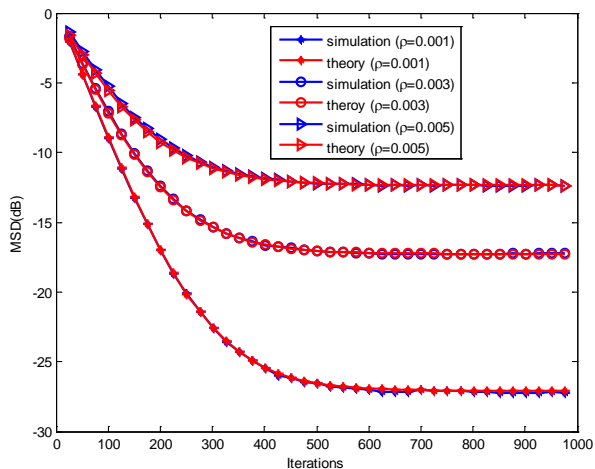

Fig. 12 The experimental and theoretical MSD curves of the proposed ATC-LRZA-DLMS algorithm with respect to $\rho$.

Finally, we implement the verification for the ATC-LRZA-DLMS algorithm with respect to $\rho$, shown in Fig. 12. The parameters are set to $\mu = 0.008$, $\gamma = 0.001$, $\varepsilon = 1$. Similar to the results in the ATC-LZA-DLMS algorithm, the parameter $\rho$ also has a great impact on the performance of the ATC-LRZA-DLMS algorithm.

### D. STEADY STATE THEORETICAL VALIDATION

We here evaluate the steady state theoretical analysis for the ATC-LZA-DLMS and ATC-LRZA-DLMS algorithms. The network and the unknown system are consistent with that in the transient validation. The steady state MSD curves are obtained from (48) to (50). As can be seen from Figs. 13 and 14, the steady state theoretical values are in good agreement with the experimental results.

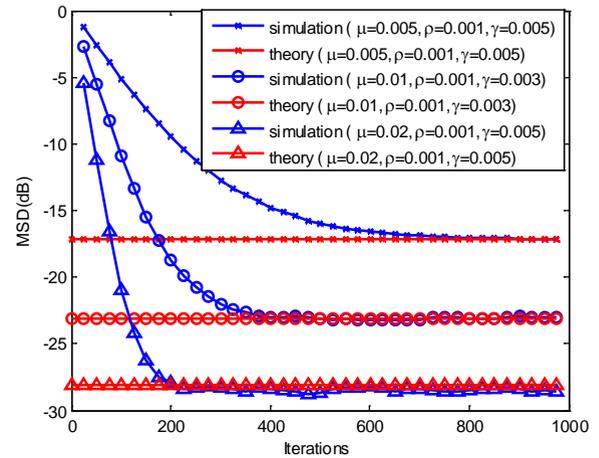

Fig. 13 The experimental and theoretical MSD curves of the proposed ATC-LZA-DLMS algorithm.

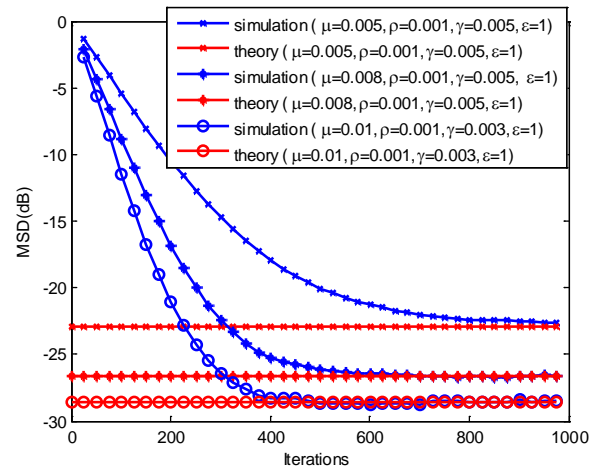

Fig. 14 The experimental and theoretical MSD curves of the proposed ATC-LRZA-DLMS algorithm.

### V. CONCLUSIONS

In this paper, by incorporating the zero-attractors into the leaky DLMS algorithm, we have proposed the LZA-DLMS and LRZA-DLMS algorithms including their ATC and CTA versions. For sparse system identification, the proposed algorithms outperform various existing algorithms when the inputs are colored. In particular, in the case of time-varying sparse system, the LRZA-DLMS algorithms exhibit superior performance than the LZA-DLMS algorithms thanks to the reweighted regularization. Employing several common assumptions and approximations, we have achieved the theoretical recursion, which can successfully characterize the transient network MSD of our findings. To guarantee the convergence in the mean and mean-square, the stability bound of the step-size for the proposed ATC-type algorithms has been determined. Moreover, we have implemented the steady state theoretical analysis for the ATC-type algorithms. Extensive experiments under various conditions have shown the MSD curves of theoretical analysis match accurately with the experimental curves. In our future work, we will carry out the research on the time-varying leaky factor to enhance estimation performance.

**APPENDIX A**

After a simple matrix calculation for (36), the steady state value of $\text{Tr}(\tilde{\mathbf{W}}_\infty)$ is given by

$$\text{Tr}(\tilde{\mathbf{W}}_\infty) = \text{Tr}(\text{vec}^{-1}(\boldsymbol{\Phi}(\mathbf{B}^T \otimes \mathbf{B}^T)\text{vec}(\mathbf{W}_o)))$$
$$+ \text{Tr}(\text{vec}^{-1}(\boldsymbol{\Phi}(\mathbf{B}^T \otimes \mathbf{B}^T)\text{vec}(E(\mathbf{w}_o \tilde{\mathbf{w}}_\infty^T))))$$
$$+ \text{Tr}(\text{vec}^{-1}(\boldsymbol{\Phi}(\mathbf{B}^T \otimes \mathbf{B}^T)\text{vec}(E(\tilde{\mathbf{w}}_\infty \mathbf{w}_o^T))))$$
$$+ \text{Tr}(\text{vec}^{-1}(\boldsymbol{\Phi}(\mathbf{C}^T \otimes \mathbf{C}^T)\text{vec}(E[\mathbf{U}_\infty^T \mathbf{v}_\infty \mathbf{v}_\infty^T \mathbf{U}_\infty])))$$
$$+ \text{Tr}(\text{vec}^{-1}(\boldsymbol{\Phi}(\mathbf{D}^T \otimes \mathbf{D}^T)\text{vec}(E(g[\mathbf{w}_\infty]g[\mathbf{w}_\infty^T]))))$$
$$+ \text{Tr}(\text{vec}^{-1}(\boldsymbol{\Phi}(\mathbf{B}^T \otimes \mathbf{A}^T)\text{vec}(E[\tilde{\mathbf{w}}_\infty \mathbf{w}_o^T])))$$
$$+ \text{Tr}(\text{vec}^{-1}(\boldsymbol{\Phi}(\mathbf{D}^T \otimes \mathbf{A}^T)\text{vec}(E(\tilde{\mathbf{w}}_\infty g[\mathbf{w}_\infty^T]))))$$
$$+ \text{Tr}(\text{vec}^{-1}(\boldsymbol{\Phi}(\mathbf{A}^T \otimes \mathbf{B}^T)\text{vec}(E[\mathbf{w}_o \tilde{\mathbf{w}}_\infty^T])))$$
$$+ \text{Tr}(\text{vec}^{-1}(\boldsymbol{\Phi}(\mathbf{D}^T \otimes \mathbf{B}^T)\text{vec}(E(\mathbf{w}_\infty g[\mathbf{w}_\infty^T]))))$$
$$+ \text{Tr}(\text{vec}^{-1}(\boldsymbol{\Phi}(\mathbf{A}^T \otimes \mathbf{D}^T)\text{vec}(E(g[\mathbf{w}_\infty]\tilde{\mathbf{w}}_\infty^T))))$$
$$+ \text{Tr}(\text{vec}^{-1}(\boldsymbol{\Phi}(\mathbf{B}^T \otimes \mathbf{D}^T)\text{vec}(E(g[\mathbf{w}_\infty]\mathbf{w}_\infty^T))))$$
(56)

Using the property $\text{Tr}(XY) = (\text{vec}(X))^T \text{vec}(Y)$ and letting $X = \mathbf{I}_{MN}$ and $Y = \text{vec}^{-1}(\boldsymbol{\Phi}(\mathbf{B}^T \otimes \mathbf{B}^T)\text{vec}(\mathbf{W}_o))$ for the first term on the RHS of (56), we arrive at

$$\text{Tr}(\text{vec}^{-1}(\boldsymbol{\Phi}(\mathbf{B}^T \otimes \mathbf{B}^T)\text{vec}(\mathbf{W}_o)))$$
$$= \text{vec}(\mathbf{I}_{MN})^T \boldsymbol{\Phi}(\mathbf{B}^T \otimes \mathbf{B}^T)\text{vec}(\mathbf{W}_o)$$
(57)

Performing the similar operation for the remaining terms in (56), we finally obtain (49).

**ACKNOWLEDGMENT**

This work was partially supported by National Science Foundation of P.R. China (Grant: 61571374, 61871461 and 61433011).